\documentclass[%
reprint,
showpacs,
showkeys,
amsmath,amssymb,
 aps,
]{revtex4-1}

\usepackage{float}
\usepackage{graphicx}
\usepackage{dcolumn}
\usepackage{bm}
\usepackage{color}
\usepackage{xcolor}

\begin{document}

\preprint{APS/123-QED}

\title{Speed-accuracy tradeoff and its effect in the game of cricket: predictive modeling from statistical mechanics perspective}

\author{Mohd Suhail Rizvi}
 \email{suhailr@bme.iith.ac.in}
 \affiliation{%
 Department of Biomedical Engineering, Indian Institute of Technology Hyderabad, Kandi, Telangana, 502285, India.
}%

\date{\today}

\begin{abstract}
The speed-accuracy tradeoffs are prevalent in a wide range of physical systems. In this paper, we demonstrate speed-accuracy tradeoffs in the game of cricket, where `batters' score runs on the balls bowled by the `bowlers'. It is shown that the run scoring rate by a batter and the probability of dismissal follow a power-law relation. Due to availability of extensive data, game of cricket is an excellent model for the study of the effect of speed-accuracy tradeoff on the overall performance of the system. It is shown that the exponent of the power-law governs the nature of the adaptability of the player in different conditions and can be used for their assessment. Further, it is demonstrated that the players with extreme values of the power-law exponent are better suited for different playing conditions as compared to the ones with moderate values. These findings can be utilized to identify the potential of the cricket players for different game formats and can further help team management in devising strategies for the best outcomes with a given set of players.
\end{abstract}
\maketitle

\section{Introduction} \label{sec:introduction}
In a wide variety of systems, be they natural or man-made, the operational speed and functional reliability do not go hand-in-hand and this phenomenon is known as ``speed-accuracy tradeoff'' \citep{pmid20709093,pmid18217850,pmid2969031,nilsson2004thesis,pmid25554788,pmid10096999,pmid21958757,pmid24966810,wickelgren1976}. The speed in different contexts can represent speed of the physical motion \citep{pmid18217850,nilsson2004thesis,pmid10096999}, characteristic time of decision making or memory retrieval process \citep{pmid2969031,pmid21958757}, rate of cell proliferation \citep{pmid25554788}, rate of a chemical reaction or a natural process \citep{pmid20709093,pmid24966810,wickelgren1976}, and correspondingly the functional reliability or accuracy stands for the deviation from the given target in case of physical motion, error in making right decision, chances of lethal genetic mutation or formation of undesired products. The presence of speed-accuracy tradeoffs in such diverse range of scenarios indicates towards underlying similarities among these system.

We show in this paper that similar speed-accuracy tradeoffs are also present in the game of cricket and can be utilized as the performance indicators of players. The game of cricket, arguably the second most popular sport in the world \citep{economist_article}, played between two teams, primarily involves three set of players- batters, bowlers and fielders. On a cricket pitch, a bowler of one team (`bowling' team) throws a ball towards a batter of the opponent team (`batting' team) who, in turn, hits the ball with a bat. The fielders of the bowling team try to collect or catch the ball after the batter hits it and bring it back to the pitch. Meanwhile, the goal of batters is to score points, or `runs' as called in cricket, by running between the wickets present at each end of the pitch such that they reach the wicket before a fielder can collect the ball and knock the wickets down. The batters can also score $4$ or $6$ runs on a delivery by hitting the ball outside of a marked boundary in a grounded fashion or aerially, respectively. Thus normally a batter can score $0$, $1$, $2$, $3$, $4$, $5$ or $6$ runs at each delivery. If, however, either the wicket is knocked down or the fielders take a straight catch before the hit ball touches the ground, the batter is declared ``out'' (also known as `losing wicket') and can no longer bat in that game. 
When a batter gets out, the next batter comes to bat and it goes on until all the players of the batting team are out, also known as the end of an `inning'. This is followed by the second innings where the opponent team bats to score more runs. Further, in fixed `overs' (six consecutive balls bowled) format of the game the inning of a team can also end if the prescribed overs are finished. For instance, in T20I format of the game, one inning can go on for $20$ overs (120 balls), in one-day international (`ODI') cricket, it goes for $50$ overs, and in `test' format of the cricket, there are virtually unlimited overs available for batting.

The availability of extensive statistical data about the international as well as club level cricket matches has encouranged the use of various statistics, such as strike rate (average runs scored for each ball faced), average runs scored for each inning, number of innings with a score of $50$ or $100$ for batters as the measure of their batting abilities. Similar statistics, such as economy rate (average runs conceded for each over bowled), strike rate (average runs conceded for each wicket), are also used for the performance assesment of bowlers. Although these quantities do represent the quality of the player in different formats of the game they do not capture the changes across the three formats of the game. Here, in this report, we demonstrate that the players' performances across three game formats are linked to each other in the form of speed-accuracy tradeoffs in batting as well as bowling. For this perpose we define ``speed'' in the batting (bowling) as the run scoring (conceding) rate and ``accuracy'' as the balls faced (delivered) before losing (gaining) his/her (opponent's) wicket. We also show that these tradeoffs are poswerful indicators of the players' performances and predictors across the game formats. We also utilize predictive modeling to identify the players better suited for different playing conditions. 

\section{Data acquisition and Methods} \label{sec:mat_n_meth}
For the analysis, we obtained the data from ESPN Cricinfo \citep{cricinfo} website for the international T20I, ODIs, and test matches. For each batter, the obtained data included total runs scored, total balls faced, total number of times they got out and their respective run scoring rate. In the test, ODI and T20I formats of the game, players who have scored a total of $5000$, $3000$ and $500$ runs or more, respectively, were considered for analysis. Similarly for each bowler, total runs conceded, total balls bowled and average run given per ball were obtained. The bowlers who have taken $100$, $100$ and $20$ wickets in the tests, ODIs and T20Is, respectively, were taken for the study. 

The statistical analyses were performed using package $R$ and the predictive modeling of the batting was implemented using MATLAB. For each statistical analysis, $p$-value and effect size were obtained and are reported with the data. For the predictive parameter, the exponent $\alpha$, $95\%$ confidence intervals were also calculated. For the collective analysis of all the players (Fig. \ref{fig:powerlaw} all the batters and bowlers as selected from the criteria mentioned previously were considered. For the analysis of individual players (Figs. \ref{fig:playerwise}C and \ref{fig:playerwise}D), only those players were selected for whom complete statistics were available for all three cricket formats which resulted in $24$ batters and $16$ bowlers for the analysis.

\section{Results and Discussion} \label{sec:results_discuss}
\subsection{Speed-accuracy tradeoffs in cricket}
\begin{figure*}[ht]
\centering
\includegraphics[width=0.8\linewidth]{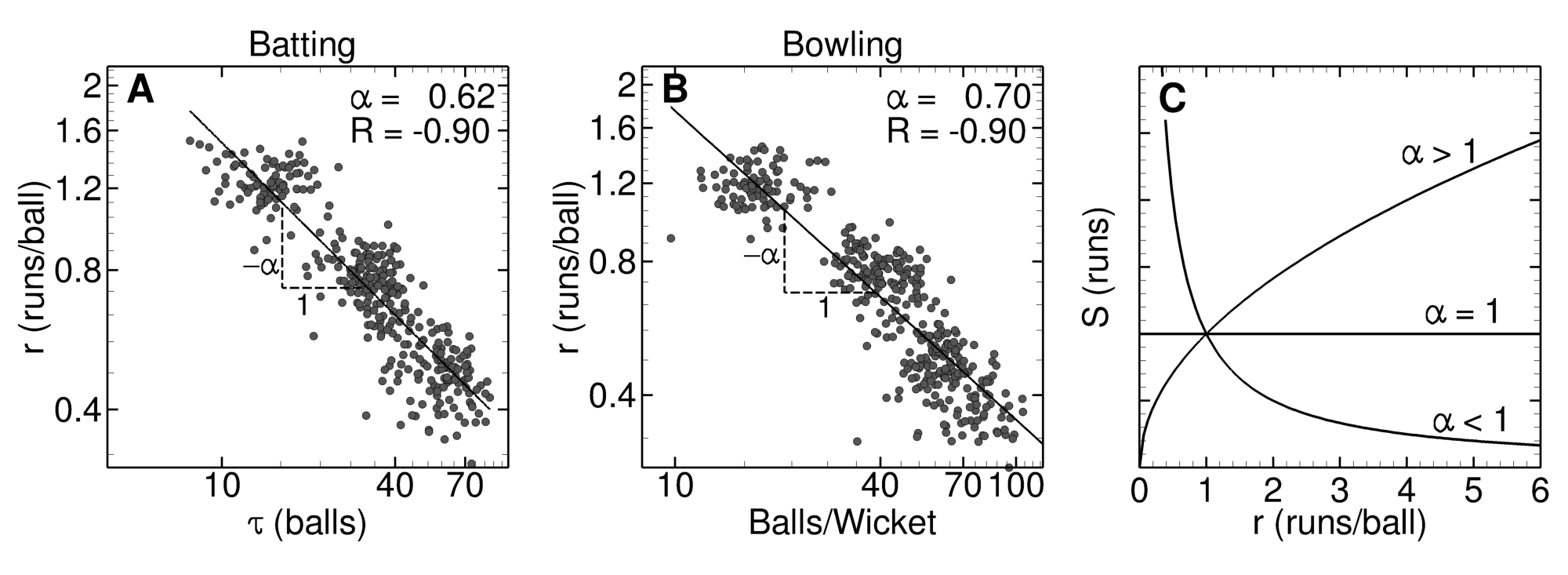}
\caption{Power law relationships in the game of cricket. (A) The run scoring rate, $r$, measured as runs scored per ball faced, and the duration of batting before getting out, quantified in terms of half life, $\tau$, for batters are related in a power law relationship. (B) For the bowlers also, the run conceding rate, $r$, and balls delivered before taking a wicket follow power law relationship. In both plots, $\alpha$ and $R$ represent the power law exponent (see relation \ref{eq:powerlaw} and the coefficient of correlation, respectively. (C) The variation in the total runs scored, $S$, in unlimited number of balls as a function of runs scoring rate, $r$, for three different $\alpha$. The runs in (C) are shown at an arbitrary unit.}
\label{fig:powerlaw}
\end{figure*}

From the data obtained from the ESPN Cricinfo \citep{cricinfo} website, we calculate two parameters for batters- the runs scoring rate, $r$, and the half-life, $\tau$, of the innings. The run scoring rate is defined as the average number of runs scored by a batter at each ball faced, whereas the average half life of a batter's inning is given by
\begin{equation}
\tau = \frac{\log 2}{p_e} = \frac{B\log 2}{d}
\end{equation}
where $p_e$ is the probability of getting out for a batter at a ball, and $B$ and $d$ are the total number of balls faced and the number of dismissals, respectively. We found that for batters the run scoring rate and the innings half-life are related through a power law relationship as shown in Fig. \ref{fig:powerlaw}A and Table 1. That is 
\begin{equation} \label{eq:powerlaw}
r = K_{bat} \tau^{-\alpha} = K_{bat} \log 2 \left( \frac{B}{d}\right)^{-\alpha}
\end{equation}

\begin{table}[!h]
\centering
\caption{Summary of the collective statistical analysis of all the players (Fig. 1).}
\begin{tabular}{llll}
\hline
Player type & $\alpha$ & Effect size (Pearson's $R$) & $p$-value \\ 
\hline
batters & $0.618$ & $-0.902$ & $<10^{-15}$ \\ 
Bowlers & $0.695$ & $-0.898$ & $<10^{-15}$ \\
\hline
\end{tabular}
\end{table}

where $K_{bat}$ is a phenomenological constant associated with the batting and $\alpha$, the power law exponent, dictates the nature of the batters's adaptability under different circumstances. Similarly for the bowling too, the runs conceding rate and the balls delivered for a wicket follow a power law relation (Fig. \ref{fig:powerlaw}B, Table 1). Owing to this relationship, the total runs, $S$, scored by a batter before getting out are always bounded as  
\begin{equation} \label{eq:runscored}
S \sim \tau r \sim r^{\left(1-1/\alpha\right)},
\end{equation}
is dependent on $\alpha$ and the run scoring rate. The relation \ref{eq:runscored} demonstrates that the exponent $\alpha$ is an indicator of the player performance with which the cricket batters can be categorized in three categories (Fig. \ref{fig:powerlaw}C). For the batters with $\alpha>1$, the total scored runs, $S$, increase with an increase in the run scoring rate, $r$, whereas for $\alpha<1$ and $\alpha=1$ the total runs decrease and remain unchanged with an increase in the run scoring rate, respectively. This shows that for an average batter ($\alpha=0.62$, Fig. \ref{fig:powerlaw}A) an increase in the run scoring rate leads to a reduction in the total runs scored. Similarly, for an average bowler ($\alpha=0.7$, Fig. \ref{fig:powerlaw}B) an increase in the run conceding rate results in lesser runs conceded as it leads to faster dismissal of opposing batters. Therefore, the exponent $\alpha$ is an indicator of the batting and bowling performances of the players in different conditions. 

\subsection{Analysis of individual players}
\begin{figure}[!ht]
\centering
\includegraphics[width=\linewidth]{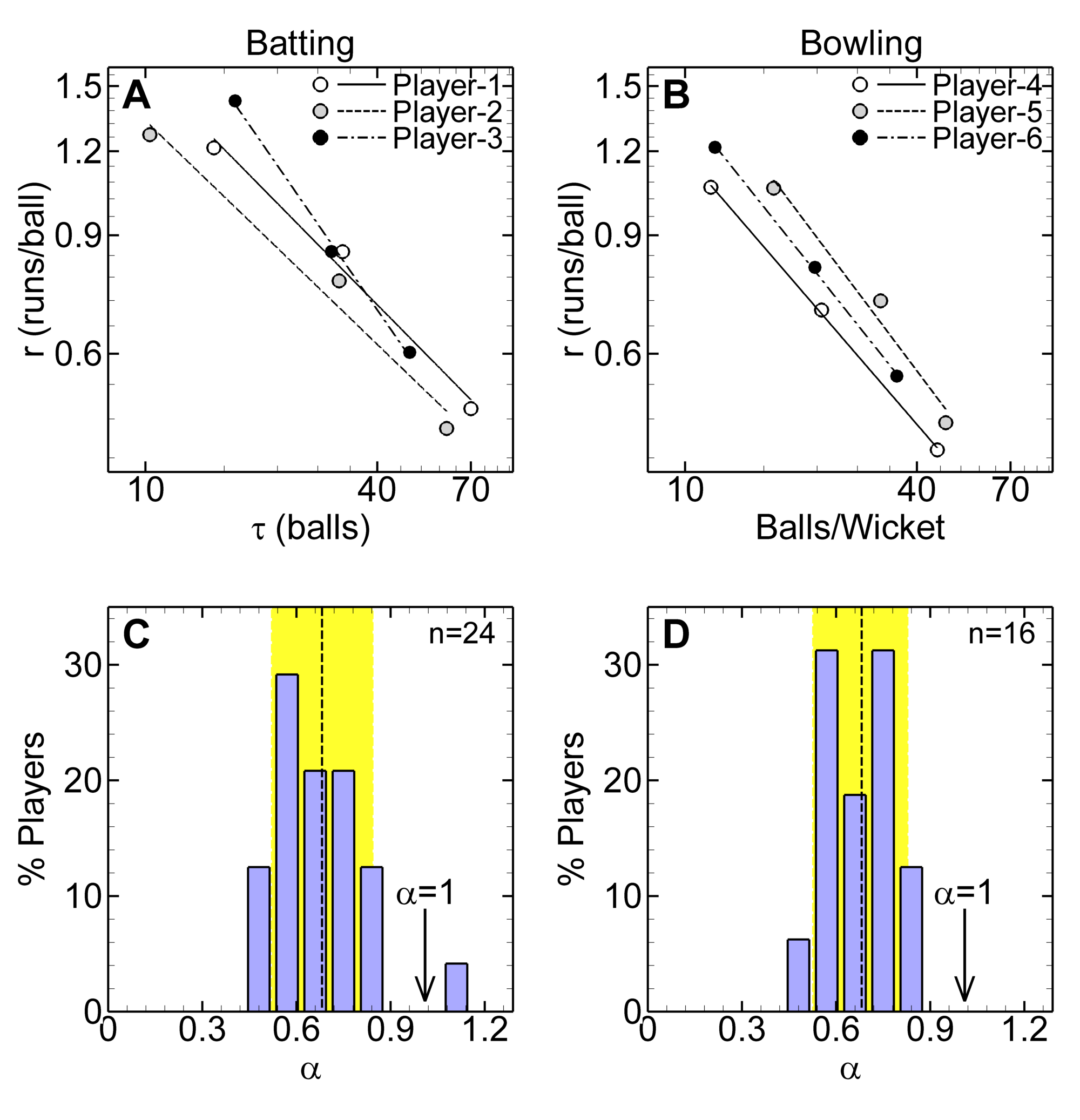}
\caption{Power law exponents for individual players. Power law fits between (A) the run scoring rate, $r$, and innings half life, $\tau$, for three representative batters, and (B) run conceding rate and number of deliveries before a wicket for three bowlers, across three cricket formats. Distributions of power law exponents, $\alpha$, for individual (C) batters and (D) bowlers. In (C) and (D) the dashed line correspond to the respective average values of $\alpha$ and the yellow region span the width equivalent to the two standard deviations of the distributions. $n$ stands for the number of players considered in (C) and (D).}
\label{fig:playerwise}
\end{figure}

In Fig. \ref{fig:powerlaw}, an average power law relationship is obtained for all the players together, but similar relation can also be obtained for the individuals. The players who have played three formats of the game of cricket (tests, ODIs and T20Is) have three sets of run scoring rate, $r$, and half life $\tau$, each. The log-log plot of the run scoring rate and half-life for individual players demonstrates that, here too, the power-law relation holds for batting (Fig. \ref{fig:playerwise}A, Table 2) as well as bowling (Fig. \ref{fig:playerwise}B, Table 2). 

\begin{table}[!h]
\centering
\caption{Summary of the statistical analysis of individual players (Fig. 2).}
\begin{tabular}{llll}
\hline
Player type & average $\alpha$ & $95\%$ CI for $\alpha$ & Effect size (Pearson's $R$)\\ 
\hline
batters & $0.680$ & $[0.617, 0.743]$ & $<-0.880$ \\ 
Bowlers & $0.679$ & $[0.623, 0.734]$ & $<-0.930$ \\
\hline
\end{tabular}
\end{table}

Therefore, from these relationships power-law exponents, $\alpha$, were  obtained for each batter and bowler. The respective frequency distributions of the $\alpha$ are shown in Figs. \ref{fig:playerwise}C and \ref{fig:playerwise}D. The frequency distributions show that $0.4 \le \alpha \le 1.1$ for batting as well as bowling (see Table 2 for $95\%$ CI). It can be seen that for a majority of players $\alpha<1$ which indicates that for an average batter ($\alpha<1$) an increase in the run scoring rate leads to a decrease in the total runs scored (Fig. \ref{fig:powerlaw}C).

As the power-law relation can also be interpreted in terms of the proportional relative change between the two quantities, that is 
\begin{equation}
\frac{\partial r}{r} = -\alpha \frac{\partial \tau}{\tau},
\end{equation}
the batters with high $\alpha$ value are more suited for the batting at higher run scoring rates. This is because an increase in their run scoring rate results in lesser proportional decrease in their innings half-life, $\tau$. By the same argument, the batters with very low $\alpha$ value are more apt at lower run scoring rates. Similarly, for bowlers, lower $\alpha$ makes the player suitable for the shorter formats of the game where higher run scoring rates are more common. 

Therefore, the exponent $\alpha$ is a powerful indicator of the players' adaptability in different situations where high or low run scoring rates are desired. This parameter can be utilized to identify the suitability of a given player for different formats of the game, such as the batters with high $\alpha$ are more suitable for shorter games (T20Is) whereas the ones with smaller $\alpha$ are more appropriate for longer formats of the game (tests).

\subsection{Predictive model for run scoring}
In order to understand the influence of the power-law relationship between $r$ and $\tau$ on the overall run scoring by a batter, we model the score evolution as a one-dimensional random walk with a drift. This model can also be applied to the bowlers with apppriate changes in the model parameters. Previously, using the available ESPN Cricinfo \citep{cricinfo} data, the anomalous diffusion nature of the score evolution in the game of cricket has been shown \citep{pmid23005806}. Although, the score evolution for the whole team follows anomalous diffusion \citep{pmid23005806}, we model the run scoring by a single batter as a normal diffusion. If $P_S\left(S,B\right)$ is the probability of scoring $S$ runs after playing $B$ balls, then the expression for $P_S\left(S,B\right)$ can be written as the following recurrence relation
\begin{equation}\label{eq:discrete_score}
P_S \left(S,B\right) = \sum \limits_{i=0}^{6} \left[ \left(1-p_e\right)p_i P_S \left(S-i,B-1\right) \right] - p_e P_S \left(S,B-1\right)
\end{equation}
where $p_i$ is the run-scoring distribution for a player, that is the probability of scoring $i$ runs in a single ball and $\displaystyle p_e=\frac{\log2}{\tau}$ is the probability for the batter of getting out at a ball. Here the two terms on the right hand side of the recurrence relation correspond to scoring runs and gertting out at a given ball, respectively. As a player can score at most $6$ runs at each ball the summation is performed for $0\le i \le 6$. For the ease of analysis the equation \ref{eq:discrete_score} can be written in the continuum limit as 
\begin{equation}
\frac{\partial P_s\left(s,b\right)}{\partial b} = -r \frac{\partial P_s\left(s,b\right)}{\partial s} + D \frac{\partial^2 P_s\left(s,b\right)}{\partial s^2} - r_e P_s\left(s,b\right)
\end{equation}
where $P_s$, $s$ and $b$ are the continuum counter-parts of discrete variables $P_S$, $S$ and $B$, respectively and $r=\left(1-p_e\right)\sum \limits_{i=0}^{6} i p_i$, $\displaystyle D=\frac{1-p_e}{2}\sum \limits_{i=0}^{6} i^2 p_i$ and $r_e=-\log \left(1-p_e\right)$\citep{feldman}. The left hand side of the equation is the change in the probability $P_s$ at each ball and the three terms on the right hand side correspond to the average scoring rate for a batter, the variation in the run scored by batter at each ball and dismissal at a given ball, respectively. It has to be noted that this predictive model is about the average performance of a player where all other factors remain unchanged. 

For this model of diffusion, the probability of reaching a target score of $s$ runs in $b$ balls or less can be obtained by the integral of the first-passage time distribution \citep{ding_ranga_2004} as
\begin{equation} \label{eq:probTarget}
\psi \left(s,b\right) = \int \limits_0^b \frac{s}{\sqrt{4\pi D t^3}} \exp \left(-r_e t - \frac{\left(s-rt\right)^2}{4Dt}\right)dt
\end{equation}
The difference between $P_s\left(s,b\right)$ and $\psi \left(s,b\right)$ has to be noted where $P_s\left(s,b\right)$ is the probability density of scoring $s$ runs in $b$ balls whereas $\psi \left(s,b\right)$ is the probability denisity of achieving $s$ runs in $b$ balls {\it or less}. With this expression, the dynamics of the score evolution for a batter can be studied under two scenarios. In the first scenario the number of balls available for a batter is fixed and it is desired to score as many runs as possible. In the second scenario, however, the target of the runs to be scored is fixed and a batter is expected to achieve that target in as few balls as possible. We study these two scenarios in the following. 

\subsubsection{Scenario 1: Runs scoring in first inning of a game}
\begin{figure*}[!ht]
\centering
\includegraphics[width=0.8\linewidth]{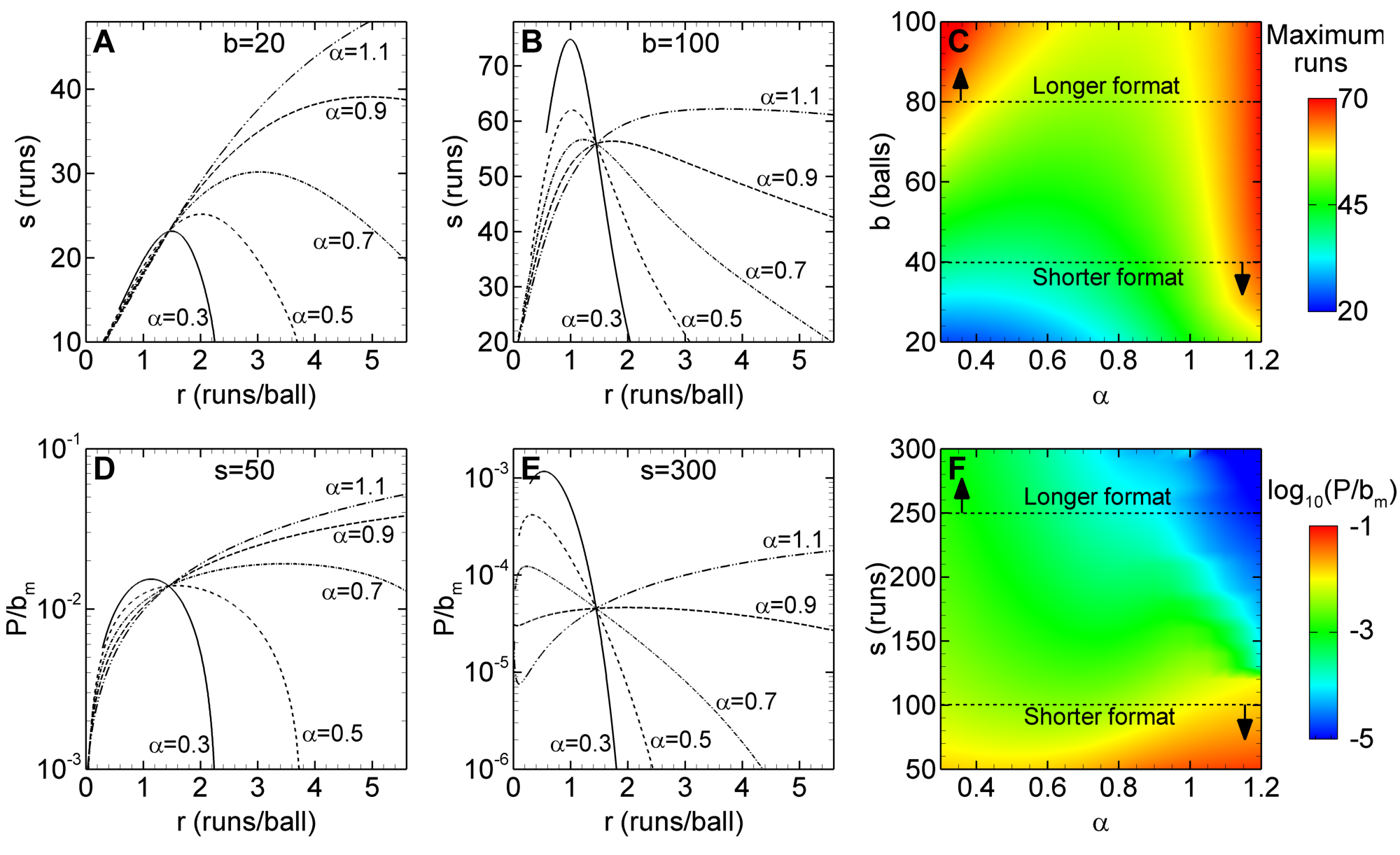}
\caption{Drift-diffusion based predictive model of run scoring and target chasing in cricket. Total runs scored, $s$, against run scoring rate, $r$, in (A) $b=20$ balls, and (B) $b=100$ balls, as obtained from the continuum diffusion model of the score evolution for different $\alpha$. (C) Maximum possible runs scored by batters with different $\alpha$ in finite number of balls. Rate of reaching a target score of (D) $s=50$ runs, and (E) $s=300$ runs as a function of $r$. (F) Highest rate of chasing a finite target for different $\alpha$. In (C) and (F), smaller number of balls and shorter target scores, respectively, stand for the shorter formats of the game of cricket (T20Is), whereas the large $b$ and $s$ correspond to longer format (tests). The arrows in (C) and (F) mark the most suitable $\alpha$ in two formats of the game.}
\label{fig:fixedBalls_fixedRuns}
\end{figure*}

In the fixed overs format of the cricket (ODIs and T20Is), in the first inning of a match the number of balls to be played by a batter are fixed. Therefore, in this scenario, the runs scored by the batter for different run scoring rates $r$ can be studied by incorporating the power-law relation between $r$ and $\tau$. By substituting 
\begin{equation}\label{eq:re_r}
r_e = -\log \left( 1 - \frac{\log2}{\tau_0} \left( \frac{r_0}{r}\right)^{-1/\alpha}\right)
\end{equation}

in relation \ref{eq:probTarget}, the total runs scored by a batter in $b$ balls can be obtained for different run scoring rates as
\begin{equation}
s_m \left(b\right) = \int \limits_0^{\infty} \int \limits_0^{b} \frac{s}{\sqrt{4\pi D t^3}} \exp \left(-r_e t - \frac{\left(s-rt\right)^2}{4Dt}\right)dtds.
\end{equation}
The relation \ref{eq:re_r} is obtained by the substitution of power-law relation in $r_e=-\log\left(1-p_e\right)$, where $r_0$ and $\tau_0$ are the reference run scoring rate and inning half-life, respectively. For two extreme cases, $b=20$ and $b=100$, the total runs scored for batters with different $\alpha$ are shown as a function of run scoring rate $r$ in Figs. \ref{fig:fixedBalls_fixedRuns}A and \ref{fig:fixedBalls_fixedRuns}B, respectively. For low $\alpha$, the runs scored in finite number of balls vary non-monotonically with the run scoring rate (Figs. \ref{fig:fixedBalls_fixedRuns}A-B). On the other hand, for high $\alpha$, the runs scored vary monotonically for smaller number of balls (Figs. \ref{fig:fixedBalls_fixedRuns}A-B). Further, these non-monotonic relationships show that for a fixed number of balls the maximum possible runs which a batter can score can be estimated. Fig. \ref{fig:fixedBalls_fixedRuns}A shows that for small number of balls the batter with higher $\alpha$ would score more runs than the batter with lower $\alpha$. On the other hand, for large number of balls, the batter with smaller $\alpha$ is predicted to score higher runs (Fig. \ref{fig:fixedBalls_fixedRuns}B). The maximum runs scored by batters for different $\alpha$ and number of balls are shown in Fig. \ref{fig:fixedBalls_fixedRuns}C. This demonstrates that in the first inning of the shorter format of the game (smaller number of balls) the batters with larger $\alpha$ would score more runs (downward arrow in Fig. \ref{fig:fixedBalls_fixedRuns}C), and in the longer format (larger number of balls) batters with smaller $\alpha$ would perform better (upward arrow in Fig. \ref{fig:fixedBalls_fixedRuns}C). This also shows that batters with $\alpha>1$, which are rare (Fig. \ref{fig:playerwise}C), always perform better irrespective of the game format.

\subsubsection*{Scenario 2: Target chasing in second inning of a game}
In the second inning of a fixed over format game of cricket, the batters are tasked to chase a target score. In chasing a target score, it is important to see if a batter with a specific $\alpha$ can achieve the target at all. The probability for a batter to reach a target score $s$ is given by
\begin{equation}\label{eq:prob_reach_target}
P\left(s\right) = \lim \limits_{b \rightarrow \infty} \psi\left(s,b\right) = \int \limits_0^{\infty} \frac{s}{\sqrt{4\pi D t^3}} \exp \left(-r_e t - \frac{\left(s-rt\right)^2}{4Dt}\right)dt,
\end{equation}
where $r_e$ is given by relation \ref{eq:re_r}. Further, in addition to the probability of reaching a target score (relation \ref{eq:prob_reach_target}, the average number of balls required to achieve the target is also an important quantity. It is given by
\begin{equation} \label{eq:averageBalls}
b_m\left(s\right) = \frac{s}{P\left(s\right)} \int \limits_0^{\infty} \frac{t}{\sqrt{4\pi D t^3}} \exp \left(-r_e t - \frac{\left(s-rt\right)^2}{4Dt}\right)dt.
\end{equation}
The quantities $P\left(s\right)$ and $b_m\left(s\right)$ can be combined as $P/b_m$ to obtain the effective rate of reaching a target score. Low values of $P/b_m$ imply that the probability of reaching the target score is low and the average number of required balls for the same is high. On the other hand, high $P/b_m$ indicate high probability and smaller number of required balls for achieving the target score. As shown in Figs. \ref{fig:fixedBalls_fixedRuns}D and \ref{fig:fixedBalls_fixedRuns}E, the quantity $P/b_m$ varies non-monotonically with the run scoring rate, $r$. Further, for smaller targets, the batter with higher $\alpha$ would chase the runs more effectively (Fig. \ref{fig:fixedBalls_fixedRuns}D), and for larger target the batter with smaller $\alpha$ are more reliable (Fig. \ref{fig:fixedBalls_fixedRuns}E). The highest value of $P/b_m$ for different values of $\alpha$ and target scores are shown in Fig. \ref{fig:fixedBalls_fixedRuns}F, which demonstrates the suitability of batter with larger $\alpha$ for shorter run chase (downward arrow), and  batter with smaller $\alpha$ for longer run chases (upward arrow). 

The analysis of the two scenarios (run scoring in the first inning and target chasing in the second inning) demonstrates that the batters with the larger $\alpha$ are more suitable for the shorter format of the game (T20Is), whereas batters for smaller $\alpha$ are more apt for longer format of the games of cricket (tests).

It has to be highlighted that the drift-diffusion based predictive model is based on an assumption that all the balls faced by the batters are same and the effect of different bowlers are ignored. Therefore, the predictions of the present model are for the average performance of a batter. In order to assess the batter's performace against specific bowlers similar statistical data of the particular batter is required and once it is available the same analysis is straight forward. Therefore, the statistical analysis and the predictive model presented in this paper can be used by the management of the cricket teams for the analysis of the players' performances and in identifying best strategies agaisnt any given opponent for the best outcome. 

\bibliographystyle{unsrt}
\bibliography{cricket_analytics}

\begin{thebibliography}{10}

\bibitem{pmid20709093}
C.~C. Wu, O.~S. Kwon, and E.~Kowler.
\newblock {{F}itts's {L}aw and speed/accuracy trade-offs during sequences of
  saccades: {I}mplications for strategies of saccadic planning}.
\newblock {\em Vision Res.}, 50(21):2142--2157, Oct 2010.

\bibitem{pmid18217850}
M.~Dean, S.~W. Wu, and L.~T. Maloney.
\newblock {{T}rading off speed and accuracy in rapid, goal-directed movements}.
\newblock {\em J Vis}, 7(5):1--12, 2007.

\bibitem{pmid2969031}
S.~Yantis and D.~E. Meyer.
\newblock {{D}ynamics of activation in semantic and episodic memory}.
\newblock {\em J Exp Psychol Gen}, 117(2):130--147, Jun 1988.

\bibitem{nilsson2004thesis}
G.~Nilsson.
\newblock {\em {{T}raffic safety dimensions and the power model to describe the
  effect of speed on safety}}.
\newblock PhD thesis, Lund Institute of Technology, 2004.

\bibitem{pmid25554788}
C.~Tomasetti and B.~Vogelstein.
\newblock {{C}ancer etiology. {V}ariation in cancer risk among tissues can be
  explained by the number of stem cell divisions}.
\newblock {\em Science}, 347(6217):78--81, Jan 2015.

\bibitem{pmid10096999}
R.~Plamondon and A.~M. Alimi.
\newblock {{S}peed/accuracy trade-offs in target-directed movements}.
\newblock {\em Behav Brain Sci}, 20(2):279--303, Jun 1997.

\bibitem{pmid21958757}
C.~C. Liu and T.~Watanabe.
\newblock {{A}ccounting for speed-accuracy tradeoff in perceptual learning}.
\newblock {\em Vision Res.}, 61:107--114, May 2012.

\bibitem{pmid24966810}
R.~P. Heitz.
\newblock {{T}he speed-accuracy tradeoff: history, physiology, methodology, and
  behavior}.
\newblock {\em Front Neurosci}, 8:150, 2014.

\bibitem{wickelgren1976}
W.~A. Wickelgren.
\newblock {{S}peed-accuracy tradeoff and information processing dynamics}.
\newblock {\em Acta Psychologica}, 41:67--85, 1977.

\bibitem{economist_article}
The Economist.
\newblock {A}nd the silver goes to..., 2011.

\bibitem{cricinfo}
ESPN Cricinfo.
\newblock {ESPN C}ricinfo-{S}tatsguru, 2016.

\bibitem{pmid23005806}
H.~V. Ribeiro, S.~Mukherjee, and X.~H. Zeng.
\newblock {{A}nomalous diffusion and long-range correlations in the score
  evolution of the game of cricket}.
\newblock {\em Phys Rev E Stat Nonlin Soft Matter Phys}, 86(2 Pt 1):022102, Aug
  2012.

\bibitem{feldman}
R.~M. Feldman and C.~Valdez-Flores.
\newblock {\em {{A}pplied Probability and Stochastic Processes}}.
\newblock Springer Heidelberg Dordrecht London New York, 2010.

\bibitem{ding_ranga_2004}
M.~Ding and G.~Rangarajan.
\newblock {F}irst passage time problem: A fokker-planck approach.
\newblock In L.~Wille, editor, {\em New Directions in Statistical Physics},
  pages 31--46. Springer Berlin Heidelberg, 2004.

\end{thebibliography}


\end{document}